\documentclass[preprint,showpacs,preprintnumbers,amsmath,amssymb]{revtex4}

\usepackage{graphicx}
\usepackage{dcolumn}
\usepackage{bm}

\begin{document}

\def\be{\begin{equation}}
\def\ee{\end{equation}}
\def\ba{\begin{eqnarray}}
\def\ea{\end{eqnarray}}

\preprint{ } \vskip .5in
\title{ Stabilizing radion and dilaton with brane gas and flux}

\author{Jin Young Kim\footnote{Electronic address:
jykim@kunsan.ac.kr}}
\address{Department of Physics, Kunsan National University,
Kunsan 573-701, Korea}
\date{\today}

\begin{abstract}

We consider the dynamics of moduli fields in brane gas cosmology.
By calculating the effective potentials from brane gas and bulk RR
field, we found an attractor behavior that can fix both the radion
and the dilaton. The potentials for radion and dilaton show global
minima that can provide the stabilizing forces so that they can be
stabilized dynamically. The effective potential for the
three-dimensional volume is runaway-type so that it can inflate.

\end{abstract}



\maketitle

\section{Introduction}

The primary goal of string cosmology is to find a successful
compactification that can produce inflation consistent with
current observations. The physical quantities characterizing the
shape or the size of extra dimensions are called moduli. In the
point of four dimensional theory, the existence of moduli appears
as scalar fields. They play important roles in supergravity models
obtained as low energy limit of superstring theory. For example,
since the vacuum expectation value of the dilaton determines the
strength of the gravitational coupling, a rolling dilaton implies
a changing gravitational constant. Also the extra dimensions
should be compact and small through the cosmological evolution of
the early universe. Thus the volume modulus of the extra
dimensions (radion) should be stabilized in any cosmological
models in order not to conflict with current observations. To
stabilize the moduli fields in string theory, various models and
mechanisms were introduced (see \cite{silverstein} and references
therein).

One way to achieve the moduli stabilization is adopting the
framework of string/brane gas cosmology based on the mechanism of
Brandenberger and Vafa \cite{bv,abe}. In the study of string
cosmology it is found that a gas of extended objects (strings and
branes) could affect the evolution of the universe in the presence
of nontrivial topology. A gas of strings or branes can be coupled
to a cosmological background in the same way as a gas of point
particles is coupled to a background of standard Einstein gravity.
Many of the works on string/brane gas cosmology were concentrated
on the stabilization of the radion assuming that the dilaton can
be stabilized by some other mechanism
\cite{wb0307044,pb,bw0403075,watson,cwb,campos, rador0504047,
patil, kaya,eastro,chatrabhuti,kim0608131,akk, kim08040073}.

Some of the key factors that can specify the cosmological
evolution in string cosmology are the spacetime dimensionality,
the geometry and topology, the location of sources, orientifold
planes, and fluxes. In the previous work \cite{kim08040073}, we
evaluated the effective potentials, induced by brane gas, bulk
flux, and supergravity particles, that govern the sizes of the
observed three and extra six dimensions. We found that the
internal volume can oscillate between two turning points or sit at
the minimum of the potential while the three-dimensional volume
expands indefinitely. However, in string or brane  cosmology, one
has to study the dynamics involving both the radion and the
dilaton \cite{bc,bbc,bbem,ks,cw,rador0701029,sano} since branes in
general couple to both the dilaton and the radion. Recently it has
been studied that a nonperturbatively generated potential by
gaugino condensation and a gas of strings can stabilize both the
dilaton and the radial modulus in heterotic string theory
\cite{dfb}.

The motivation of this paper is to study how the running of the
dilaton affects the behavior of the three-dimensional volume
factor and the radion. It is already known that the bulk fluxes
can generate potentials to stabilize the moduli in string theory
\cite{drs,gkp} and in Randall-Sundrum scenario \cite{gw}. We will
make use of a bulk Ramond-Ramond (RR) flux to stabilize
simultaneously both the dilaton and the radion in brane gas
formalism of type II string theory. We will show that it is
possible to stabilize both moduli dynamically by an attractor
mechanism, while the three-dimensional volume can expand
monotonically. The crucial point is that the coupling of dilaton
to brane and RR gauge field induces confining potentials for
stabilization.

The paper is organized as follows. In Sec. II, we consider a
cosmological formalism based on the dilaton gravity with a
four-form bulk RR flux and a six-dimensional brane gas. In Sec.
III, we set up the equations of motion for the dilaton, the
six-dimensional internal subspace, and the three-dimensional
subspace. Then we reduce them to motions of a particle under an
effective potential. We will show that the radion and the dilaton
can be stabilized separately if one of the two is fixed. In Sec.
IV, we consider the dynamical stabilization of the two fields. In
Sec. V, we conclude and discuss.

\section{Equations of motion}

In the point of string theory, the gravitational interaction is
described by the coupled system of metric and dilaton. We consider
the dilaton cosmology with a bulk RR flux and a brane gas. To be
specific, let us start from the following bulk effective action of
type II string theory \cite{kim0608131}
 \be
S_{\rm b} =\frac{1}{2\kappa^2} \int d^{D+1}x \sqrt{-g} \Bigl[
e^{-2 \phi} \{ R + 4 (\nabla \phi)^2 \} - \frac{1}{2 \cdot 4!}
F_4^2 - V(\phi) \Bigr], \label{bulkea}
 \ee
where $\kappa^2 = 1/ M^{D-1}_*$, with $M_*$ being the
$(D+1)$-dimensional unification scale, $\phi$ is the dilaton
field, $F_4$ is the four-form RR field strength, and $V(\phi)$ is
the dilaton potential. Although we will consider the $D=9$ case,
we will keep $D$ to see the dimensional dependence of our
analysis.

We suppose that the matter contribution of a single brane to the
action is represented by the Dirac-Born-Infeld (DBI) action of a
$p$-brane
 \be
S_{\rm p-brane} = - T_p \int d^{p+1} \xi e^{-\phi} \sqrt{ - {\rm
det} ( {\hat g}_{\mu\nu} + {\hat B}_{\mu\nu} + 2 \pi \alpha^\prime
{F}_{\mu\nu} ) } , \label{pbraneea}
 \ee
where $T_p$ is the tension of the $p$-brane and ${\hat
g}_{\mu\nu}$ is the induced metric to the brane
 \be
 {\hat g}_{\mu\nu} = g_{MN} \frac{\partial x^M}{ \partial \xi^\mu}
\frac{\partial x^N}{ \partial \xi^\nu}.
 \ee
Here $M$ and $N$ are the indices of bulk spacetime and $\mu$ and
$\nu$ are those of brane. ${\hat B}_{\mu\nu}$ is the induced
antisymmetric tensor field and ${F}_{\mu\nu}$ is the field
strength tensor of gauge field $A_\mu$ living on the brane. The
fluctuations of the fields within the brane are negligible when
the temperature is low enough. Thus we neglect ${\hat B}_{\mu\nu}$
and ${F}_{\mu\nu}$ terms. Since the DBI action couples to both the
dilaton and the radion, the brane action not only acts as a source
term for gravity but also provides potentials for dilaton and
radion.

We consider the case that the RR flux is laid on the three
dimensions where windings of brane are removed completely, while
the remaining six dimensions are wrapped with gases of branes
whose dimensions are less than or equal to $(D-3)$. In general,
the action of these gases can be written as a sum over the
contributions of the gas from each string or brane state. The
individual contribution is obtained by its number density and
energy in a hydrodynamical approximation. However, assuming that
each type of brane gas makes a comparable contribution, we
consider a gas of effective $(D-3)$-branes whose tension we denote
by $T_{D-3}$. In other words, we consider the effect of all
lower-dimensional ($ \le D-3$) brane gas as a $(D-3)$-brane gas.
The total action can be written as
 \be
 S_{\rm tot} = S_{\rm b} + S_{\rm (D-3)-brane}.
 \ee

Since the presence of moduli dependence in the Einstein term can
cause an extra tadpole for moduli from the curvature, it is
convenient to rescale the metric so that the Einstein term $\int
\sqrt{g} {\cal R}$ is completely decoupled from other moduli. To
solve the field equations we work in the ten-dimensional Einstein
frame, defined by
 \be
 g_{MN}^{\rm S} = e^{\frac{1}{2} \phi} g_{MN}^{\rm E}.
 \ee
In terms of the Einstein metric, the action can be written as
 \be
S_{\rm b}^E =\frac{1}{2\kappa^2} \int d^{D+1}x \sqrt{-g^E} \Bigl[
\{ R -\frac{1}{2} (\nabla \phi)^2 \} - \frac{e^{ \frac{1}{2} \phi
} }{2 \cdot 4!} F_4^2 - V_E(\phi) \Bigr], \label{bulkeaEin}
 \ee
 \be
S_{\rm (D-3)-brane}^E = - T_{D-3} \int d^{D-2} \xi e^{ \frac{3}{4}
\phi } \sqrt{ - {\rm det} {\hat g}_{\mu\nu}^E } ,
\label{pbraneeaEin}
 \ee
where $V_E = e^{\frac{5}{2}\phi} V (\phi)$. We drop the
superscript $E$ from now on. If one solves the field equation for
$F_4$ in the Einstein frame, the $F_4^2$ term in Eq.
(\ref{bulkeaEin}) can give potential terms for dilaton and radion.

In the point of bulk theory, the energy momentum tensor of a
single $(D-3)$-brane has a delta function singularity at the
position of the brane along the transverse directions
 \be
 \sqrt{-g} T_{\rm (D-3)-brane}^{MN} = - T_{D-3}
 \int d^{D-2} \xi e^{ \frac{3}{4} \phi } \sqrt{ - {\hat g} }
 {\hat g}^{\mu\nu} \partial_\mu x^M \partial_\nu x^N \delta
 (x - x(\xi)) .
 \ee
For cosmological setting, it seems natural to take a gas of such
branes in a continuum approximation and this smooths the
singularity by integrating over the transverse dimensions.

Since the dilaton potential is not known in string theory, we
consider the simple case where the dilaton potential in the
Einstein frame is a constant $V_E (\phi)=\Lambda$ to incorporate
the cosmological constant. We will not consider its origin in
detail here. With the metric ansatz
 \be
 ds^2 = -n^2 dt^2 + \sum_{k=1}^D a^2_k(t) (dx^k)^2 , \nonumber
 \ee
the induced metric on the brane is
 \be
 {\hat g}_{\mu\nu} = {\rm diag} \{ -(1- v^2 ), a_1^2 , a_2^2 ,
a_3^2  \} ,
 \ee
 where $v^2 = \sum_{i = 4}^{D} {\dot x}^i {\dot x}_i \ge 0$.
If we consider the static brane ($v=0$), i.e., consider the brane
does not move in the transverse directions, the brane time is the
same as the bulk time.

Taking the following form of RR field
 \be
 F_{N_1 , \cdots , N_4} = n \sqrt{4} \epsilon_{N_1 , \cdots , N_{3} }
  \nabla_0 A (t) , \label{Fansatz}
 \ee
 where $N_1 , \cdots , N_{3} \ne 0 $, the Bianchi identity,
 \be
  \nabla_{[N} F_{N_1 , \cdots , N_4 ]} = 0 , \label{Bianchi}
 \ee
is automatically satisfied since $A(t) $ is a function of $t$
only. Then the one-dimensional effective action for the scale
factors and the dilaton can be written as
 \ba
S_{\rm eff} = &-& {\cal V}_D \int dt ~n V_D \Bigl[
\frac{1}{2\kappa^2} \Bigl\{ \Bigl(\sum_{j=1}^D \frac{ \dot a_j}{
na_j}\Bigr)^2 - \sum^D_{j=1} \Bigl(\frac{\dot a_j}{na_j}\Bigr)^2
-\frac{1}{2} \frac{{\dot \phi}^2}{n^2} \Bigr\}
\nonumber \\
 &+& \Lambda +\frac{T_{D-3}}{V_3} e^{\frac{3}{4} \phi}
 - \frac{1}{2} \frac{ e^{\frac{1}{2} \phi} }{V_3^2}
 {\dot A}^2
 \Bigr], \label{effactdil}
 \ea
where ${\cal V}_D$ is the coordinate spatial volume, $V_D =
\prod_{k=1}^{D} a_k $, and $V_3 = \prod_{k=1}^{3} a_k $

To get the equations of motion, vary this action with respect to
$n$, $a_k$, $\phi$, and ${\dot A}$, then set $n=1$ at the end.
After some straightforward algebra, we obtain the following set of
equations of motion:
  \be
 \frac1{2\kappa^2} \Bigl\{ \Bigl(\sum_{j=1}^D \frac{\dot
a_j}{a_j}\Bigr)^2 - \sum_{j=1}^D \Bigl(\frac{\dot
a_j}{a_j}\Bigr)^2 - \frac{1}{2} {\dot \phi}^2 \Bigr\}
 = \Lambda +T_{D-3} \frac{e^{\frac{3}{4} \phi} }{V_3}
 + \frac{1}{2} \frac{e^{\frac{1}{2} \phi} }{V_3^2} {\dot A}^2 ,
 \label{eomn}
  \ee
  \ba
 && \frac{\ddot a_n}{a_n} + \sum_{l\ne n} \frac{\dot a_n \dot
 a_l}{a_n a_l} =  \frac{\kappa^2}{D-1} \Bigl\{ 2 \Lambda
  + (D-2) T_{D-3} \frac{e^{\frac{3}{4} \phi}}{V_3}
  + 3 \frac{e^{\frac{1}{2} \phi} }{V_3^2} {\dot A}^2
  \Bigr\}           \nonumber \\
 && - \kappa^2 \Bigl\{
  T_{D-3} \frac{e^{\frac{3}{4} \phi}}{V_D} a_n
  \frac{\partial V_{D-3}}{\partial a_n}
  -\frac{1}{2} e^{\frac{1}{2} \phi} {\dot A}^2
  a_n \frac{\partial }{\partial a_n} \Bigl( \frac{1}{V_3^2} \Bigr)
   \Bigr\} ,  \label{eoman}
  \ea
  \be
 {\ddot \phi} + \Bigl( \sum_{j=1}^D \frac{\dot a_j}{a_j}\Bigr) \dot \phi
 + 2 \kappa^2 \Bigl(
 \frac{3}{4} T_{D-3} \frac{e^{\frac{3}{4} \phi}}{V_3}
  - \frac{1}{4} \frac{e^{\frac{1}{2} \phi} }{V_3^2} {\dot A}^2
  \Bigr) = 0,      \label{eomdil}
  \ee
  \be
 \frac{d}{dt} \Bigl( V_D
 \frac{e^{\frac{1}{2} \phi} }{V_3^2} {\dot A}
 \Bigr) = 0 .  \label{adoteq}
  \ee
Note that the derivative operator $a_n \frac{\partial}{\partial
a_n}$ behaves like a counting operator. When it acts on $V_3$, it
gives 1 for $n=1,2,3$ and 0 for $n=4, \cdots , D$.

The solution of (\ref{adoteq}) is given by
 \be
 {\dot A} = Q \frac{V_3^2}{V_D} e^{- \frac{1}{2}\phi}  , \label{soladot}
 \ee
with an integration constant $Q$. Substituting (\ref{soladot}),
the equations of motion (\ref{eomn})-(\ref{eomdil}) can be written
as
 \be
 \frac1{2\kappa^2} \Bigl\{ \Bigl(\sum_{j=1}^D \frac{\dot
a_j}{a_j} \Bigr)^2 - \sum_{j=1}^D \Bigl( \frac{\dot a_j}{a_j}
\Bigr)^2 - \frac{1}{2} {\dot \phi}^2 \Bigr\}
 = \Lambda + T_{D-3} \frac{e^{\frac{3}{4} \phi} }{V_3}
 + \frac{1}{2} Q^2 \frac{e^{-\frac{1}{2} \phi} }{V_{D-3}^2} ,
 \label{eomn2}
 \ee
 \ba
 && \frac{\ddot a_n}{a_n} + \sum_{l\ne n} \frac{\dot a_n \dot
 a_l}{a_n a_l} =  \frac{\kappa^2}{D-1} \Bigl\{ 2 \Lambda
  + (D-2) T_{D-3} \frac{e^{\frac{3}{4} \phi}}{V_3}
  + 3 Q^2 \frac{e^{-\frac{1}{2} \phi} }{V_{D-3}^2}
  \Bigr\}           \nonumber \\
 && - \kappa^2 \Bigl\{
  T_{D-3} \frac{e^{\frac{3}{4} \phi}}{V_D} a_n
  \frac{\partial V_{D-3}}{\partial a_n}
  -\frac{1}{2} Q^2 e^{-\frac{1}{2} \phi}  \frac{V_3^2}{V_{D-3}^2}
  a_n \frac{\partial }{\partial a_n} \Bigl( \frac{1}{V_3^2} \Bigr)
   \Bigr\} ,  \label{eoman2}
 \ea
 \be
 {\ddot \phi} + \Bigl( \sum_{j=1}^D \frac{\dot a_j}{a_j} \Bigr) \dot \phi
 + 2 \kappa^2 \Bigl(
 \frac{3}{4} T_{D-3} \frac{e^{\frac{3}{4} \phi}}{V_3}
  - \frac{1}{4}Q^2 \frac{e^{-\frac{1}{2} \phi} }{V_{D-3}^2}
  \Bigr) = 0.      \label{eomdil2}
  \ee
To simplify the analysis we assume that the three-dimensional
space and the internal $(D-3)$-dimensional space are isotropic so
that $a_1 = \cdots = a_3 = a$ and  $a_4 = \cdots = a_D = b$. With
this assumption, we have
 \ba
 &&6\Bigl(\frac{\dot a}{a}\Bigr)^2 + (D-3)(D-4) \Bigl(\frac{\dot b}{b}\Bigr)^2 +
6(D-3) \frac{\dot a \dot b}{ab} - \frac{1}{2} {\dot \phi}^2
 \nonumber \\
&&= 2 \kappa^2 \Bigl\{ \Lambda + T_{D-3} \frac{e^{\frac{3}{4}
\phi} }{a^3}
 + \frac{1}{2} Q^2 \frac{e^{-\frac{1}{2} \phi} }{b^{2(D-3)}}
 \Bigr\},    \label{eomcons}
 \ea
 \ba
 &&\frac{\ddot a}{a} + 2\Bigl(\frac{\dot a}{a}\Bigr)^2 + (D-3)
 \frac{\dot a \dot b}{a b}  \nonumber     \\
 &&=  \frac{2 \kappa^2}{D-1} \Lambda
 + \frac{D-2}{D-1} \kappa^2 T_{D-3} \frac{e^{\frac{3}{4} \phi} }{a^3}
 - \frac{D-4}{D-1} \kappa^2 Q^2 \frac{e^{-\frac{1}{2} \phi} } {b^{2(D-3)}},
 \label{eoma}
 \ea
 \ba
 && \frac{\ddot b}{b} + (D-4)(\frac{\dot b}{b})^2 +
3 \frac{\dot a \dot b}{a b}  \nonumber  \\
&&=  \frac{2 \kappa^2}{D-1} \Lambda
 - \frac{\kappa^2}{D-1} T_{D-3} \frac{e^{\frac{3}{4} \phi} }{a^3}
 + \frac{3}{D-1} \kappa^2 Q^2 \frac{e^{-\frac{1}{2} \phi} } {b^{2(D-3)}},
 \label{eomdb}
 \ea
 \be
  {\ddot \phi} +  \Big\{ 3\frac{\dot a}{a} + (D-3) \frac{\dot b}{b} \Big\} \dot \phi
 = - 2 \kappa^2 \Bigl\{
 \frac{3}{4} T_{D-3} \frac{e^{\frac{3}{4} \phi}}{a^3}
  - \frac{1}{4}Q^2 \frac{e^{-\frac{1}{2} \phi} }{b^{2(D-3)}}
  \Bigr\}.
 \label{eomdilaton}
 \ee
Sources of positive energy in string cosmology tend to yield a
runaway behavior, driving the volume to expand. A stable or
metastable point is possible only in the presence of
counterbalancing forces. We will search the possible solutions
where $a$ becomes large indefinitely while both $b$ and $\phi$
remain finite.

\section{Effective potentials }

For simplicity, we set $2 \kappa^2 = 1$ because this factor can be
absorbed into the redefinitions of $\Lambda$, $T_{D-3}$, and
$Q^2$. Defining the two volumes of $SO(3)$ and $SO(D-3)$ subspaces
as
 \be
 \zeta \equiv a^3 ,~~~~ \xi \equiv b^{D-3},
 \ee
the equations of motion can be written as
 \be
 \frac{2}{3}\Bigl(\frac{\dot \zeta}{\zeta}\Bigr)^2 + \frac{D-4}{D-3}
 \Bigl(\frac{\dot \xi}{\xi}\Bigr)^2 +
2 \frac{\dot \zeta \dot \xi}{\zeta\xi} - \frac{1}{2} {\dot \phi}^2
=  \Lambda + T_{D-3} \frac{e^{\frac{3}{4} \phi} }{\zeta}
 + \frac{1}{2} Q^2 \frac{e^{-\frac{1}{2} \phi} }{\xi^2}
 ,    \label{conszetaxi}
 \ee
 \be
 \frac{\ddot \zeta}{\zeta} + \frac{\dot \zeta \dot \xi}{\zeta \xi}
 =3 \Bigl\{ \frac{\Lambda}{D-1}
 + \frac{D-2}{2(D-1)} T_{D-3} \frac{e^{\frac{3}{4} \phi} }{\zeta}
 - \frac{D-4}{2(D-1)} Q^2 \frac{e^{-\frac{1}{2} \phi} } {\xi^2} \Bigr\},
 \label{eomzeta}
 \ee
 \be
  \frac{\ddot \xi}{\xi}  + \frac{\dot \zeta \dot \xi}{\zeta \xi}
  = (D-3) \Bigl\{ \frac{\Lambda}{D-1}
 - \frac{1}{2(D-1)} T_{D-3} \frac{e^{\frac{3}{4} \phi} }{\zeta}
 + \frac{3}{2(D-1)} Q^2 \frac{e^{-\frac{1}{2} \phi} } {\xi^2} \Bigr\},
 \label{eomdxi}
 \ee
 \be
  {\ddot \phi} + \Bigl(\frac{\dot \zeta}{\zeta} + \frac{\dot \xi}{\xi}\Bigr) \dot \phi
  = - \frac{3}{4} T_{D-3} \frac{e^{\frac{3}{4} \phi}}{\zeta}
  + \frac{1}{4}Q^2 \frac{e^{-\frac{1}{2} \phi} }{\xi^2} .
 \label{dilatonzetaxi}
 \ee

The system of equations are second order nonlinear differential
equations with damping and driving terms. Solving the coupled set
of equations analytically seems almost impossible. We will search
perturbatively the possibility that the equations admit an
expanding solution for the unwrapped three dimensions while the
dilaton and the radion are stabilized. The starting point is to
see the behavior of each field assuming that the other field is
fixed. We will find the critical values for each field. By varying
the field equations around the critical values, one can study the
stability in perturbation theory. We follow the method used in
Ref. \cite{dfb} to analyze the stability of radion and dilaton.

Let us find the effective potential for dilaton and its critical
value. In terms of $\zeta$ and $\xi$, Eq. (\ref{dilatonzetaxi})
can be written as
 \be
 \frac{1}{\zeta\xi} \frac{d}{dt} (\zeta\xi \dot \phi )
 = - \frac{d V_{\rm eff} (\phi)}{d \phi},
 \ee
 where $V_{\rm eff} (\phi)$ is defined as
 \be
 V_{\rm eff} (\phi) = T_{D-3} \frac{e^{\frac{3}{4} \phi}}{\zeta}
   + \frac{1}{2} Q^2 \frac{e^{-\frac{1}{2} \phi} }{\xi^2} .
 \label{dilatoneff}
 \ee
Note that the coupling of dilaton to the brane gas induces a
potential term with positive exponential that drives the dilaton
toward the weak coupling limit $(\phi \to -\infty)$, while the
coupling to the RR flux gives a potential term with negative
exponential that drives the dilaton toward the strong coupling
limit $(\phi \to \infty)$. Thus the dilaton can be stabilized by
the balance of forces. The shape of the potential $V_{\rm eff}
(\phi)$, given by Fig. 1, shows a global minimum so that an
attractor solution is possible. If $\zeta$ and $\xi$ are fixed at
$\zeta =\zeta_0$ and $\xi = \xi_0$, the potential has the minimum
value at $\phi_0$ given by
 \be
 e^{\frac{5}{4}\phi_0} = \frac{\zeta_0}{3 \xi_0^2} \frac{Q^2}{T_{D-3}} .
 \label{dilatonminimum}
 \ee

\begin{figure}
\includegraphics[angle=270 , width=8cm ]{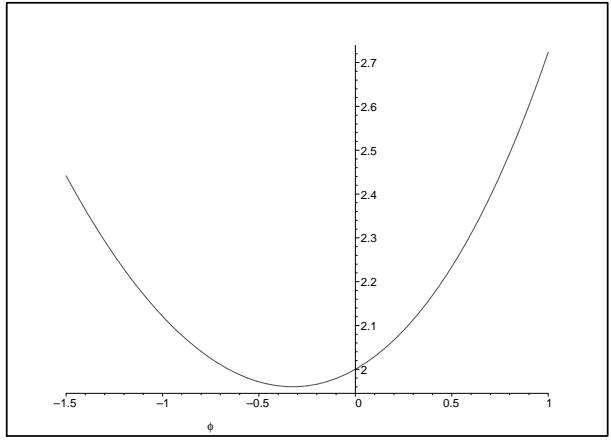} \caption{
Typical shape of the effective potential $V_{\rm eff} (\phi)$ for
the dilaton. The plot is for $\frac{1}{2} \frac{Q^2}{\xi^2}=1$ and
$\frac{T_6}{\zeta}= 1 $. } \label{fig1}
\end{figure}

\subsection{Radion stabilization for fixed dilaton}

In this subsection, assuming that the dilaton is fixed, we will
show that the radion can be stabilized while the volume of the
three-dimensional subspace grows indefinitely. When the dialton is
localized at its critical value $\phi = \phi_0$, we obtain, from
(\ref{eomzeta}) and (\ref{eomdxi}),
 \be
 \frac{\ddot \zeta}{\zeta}
 - \frac{4D-9}{2(D-1)} T_{D-3} \frac{e^{\frac{3}{4} \phi_0} }{\zeta}
 = \frac{\ddot \xi}{\xi} - \frac{D-6}{D-1}\Lambda
 - \frac{3(2D-7)}{2(D-1)} Q^2 \frac{e^{-\frac{1}{2} \phi_0} }
 {\xi^2} .
 \label{zetaxicombined}
 \ee
Since the left-hand side is a function of $\zeta$ while the
right-hand side is a function of $\xi$, we take the simplest case
by equating them to a constant $E$
 \be
 \frac{\ddot \zeta}{\zeta}
 = E + \frac{4D-9}{2(D-1)} T_{D-3}
 \frac{e^{\frac{3}{4} \phi_0} }{\zeta} ,
 \label{decoupledeomzeta}
 \ee
 \be
 \frac{\ddot \xi}{\xi} = E + \frac{D-6}{D-1}\Lambda
 + \frac{3(2D-7)}{2(D-1)} Q^2 \frac{e^{-\frac{1}{2} \phi_0} } {\xi^2} .
 \label{decoupledeomxi}
 \ee
The equations for two volume factors can be written as motions of
a particle in one-dimensional potential, ${\ddot \zeta} = -
\frac{dV_{\rm eff}(\zeta)}{d \zeta}$, ${\ddot \xi} = -
\frac{dV_{\rm eff}(\xi)}{d \xi}$, where
 \ba
 V_{\rm eff} (\zeta) &=& - \frac{1}{2} E \zeta^2
 - \frac{4D-9}{2(D-1)} e^{\frac{3}{4} \phi_0} T_{D-3} \zeta , \\
 V_{\rm eff} (\xi) &=& - \frac{1}{2} (E + \frac{D-6}{D-1} \Lambda )\xi^2
 -  \frac{3(2D-7)}{2(D-1)} Q^2 e^{-\frac{1}{2} \phi_0} \ln \xi .
  \label{radioneff}
 \ea

\begin{figure}
\includegraphics[angle=270 , width=8cm ]{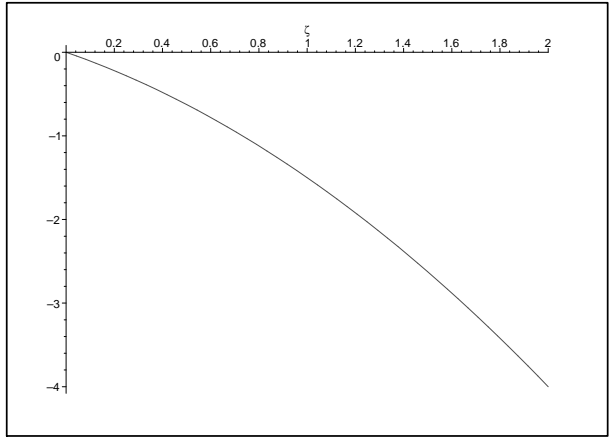} \caption{
Typical shape of the effective potential $V_{\rm eff} (\zeta)$ for
the unwrapped subvolume. The plot is for $E = 1$ and
$\frac{27}{16} T_6 e^{\frac{3}{4}\phi_0} = 1 $. } \label{fig2}
\end{figure}

\begin{figure}
\includegraphics[angle=270 , width=8cm ]{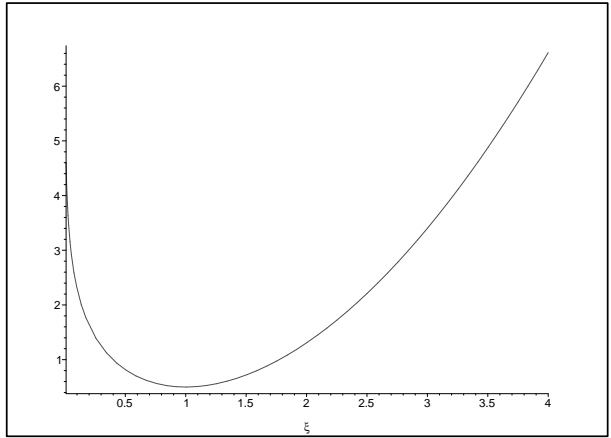} \caption{
Typical shape of the effective potential $V_{\rm eff} (\xi)$ for
the wrapped subvolume for $ \Lambda = - |\Lambda |< 0$ and $E -
\frac{3}{8} |\Lambda| < 0$. The plot is for $E - \frac{3}{8}
|\Lambda| =-1$ and $\frac{33}{16}Q^2 e^{\frac{1}{2} \phi_0} =1$.}
\label{fig3}
\end{figure}

To have a monotonic expansion for $\zeta$, we require the
condition $E>0$. In this case the shape of $V_{\rm eff}(\zeta)$ is
runaway-type as shown in Fig. 2. The solution for $\zeta$ of Eq.
(\ref{decoupledeomzeta}) is exponentially increasing $\zeta \sim
e^{\sqrt{E} t} $. Thus the three-dimensional volume can grow
indefinitely.

For the confining behavior for large $\xi$, we require the
condition
 \be
 \Lambda = - |\Lambda | < 0, ~~~~
  E - \frac{D-6}{D-1} |\Lambda| < 0.
  \label{rangeofE}
 \ee
Then the shape of the potential $V_{\rm eff} (\xi)$ is well-type
as shown in Fig. 3. Substituting $D=9$, the range of the parameter
$E$ is $0 < E < \frac{3}{8}|\Lambda| $. For parameters satisfying
(\ref{rangeofE}) the internal volume can oscillate around the
minimum of the potential or sit at the minimum of the potential
 \be
 \xi_{\rm min} = \Bigl(
 \frac{ \frac{33}{16} Q^2 e^{-\frac{1}{2} \phi_0} }
      {\frac{3}{8} |\Lambda| - E }  \Bigr)^{\frac{1}{2}} .
 \label{ximin}
 \ee

The idea for the perturbative analysis starts from setting the
asymptotic value of the radion equal to its critical value
$\xi_0=\xi_{\rm min}$. Then we examine what happens when the
radion is deviated from this value. Considering the radion
perturbation up to first order
 \be
 \xi = \xi_0 + \delta \xi ,
 \ee
we obtain
 \be
 {\ddot {\delta \xi} }
 = - \Bigl( \frac{3}{8} | \Lambda | - E\Bigr) \delta \xi
   - \frac{33}{16} Q^2 \frac{e^{-\frac{1}{2} \phi_0} } {\xi_0^2} \delta
 \xi.
 \ee
Using Eq. (\ref{ximin}), the solution is easily obtained as the
following form of oscillating solution
 \be
  {\delta \xi} =
 \delta \xi_0 \sin \omega_\xi t, ~~~~
 \omega_\xi = \sqrt{ 2 \Bigl(\frac{3}{8} |\Lambda| - E \Bigr) } .
 \ee
Thus the radion can be stabilized if the dilaton is fixed at its
critical value.

\subsection{Dilaton stabilization for fixed radion}

In this subsection, assuming that the volume factor of the
three-dimensional space grows indefinitely $\zeta \sim e^{\sqrt{E}
t}$ and the radion is fixed at $\xi_0$, we will show that the
dilaton can be stabilized.
 We consider the dilaton perturbation up to first order:
 \be
 \phi = \phi_0 + \delta \phi .
 \ee
Then we obtain the following equation for the perturbed field,
from Eq. (\ref{dilatonzetaxi}),
 \be
 {\ddot {\delta \phi} }
 + \frac{\dot \zeta_0}{\zeta_0} \dot {\delta \phi}
 + \frac{1}{8} Q^2 \frac{e^{-\frac{1}{2} \phi_0} } {\xi_0^2} \delta
 \phi
 + \frac{9}{16} T_6 \frac{e^{\frac{3}{4} \phi_0} } {\zeta_0} \delta
 \phi = 0 .
 \ee
 Substituting $\zeta_0 \sim e^{\sqrt{E}t}$ and using
 Eq. (\ref{dilatonminimum}) and Eq. (\ref{ximin}), we have
 \be
 {\ddot {\delta \phi} } + \sqrt{E} \dot {\delta \phi}
 + \frac{5}{33} \Bigl(\frac{3}{8} |\Lambda| - E \Bigr) \delta \phi =0.
 \label{dilatonperturbed}
 \ee
The general solution is given by
 \be
 \delta \phi = A e^{\alpha_{+} t}  + B e^{\alpha_{-} t} ,
 \ee
where
 \be
 \alpha_{\pm} = \frac { - \sqrt{E}
 \pm \sqrt{ \frac{53}{33} E - \frac{5}{22} |\Lambda |} }
 {2} .
 \ee

There are two possible solutions depending on the sign of the
discriminant for the parameter range $0 < E < \frac{3}{8}|\Lambda|
$.

 \noindent i) for $0 < E < \frac{15}{106}|\Lambda|$,
$\alpha_{\pm}$ are complex with their real part negative. In this
case, $\delta \phi$ can have the following form of the damped
oscillation:
 \be
 \delta \phi = \delta \phi_0 e^{- \frac{\sqrt{E}}{2} t} \sin
 \omega_\phi t, ~~~~
 \omega_\phi =  \frac{1}{2} \sqrt { \frac{5}{22} |\Lambda | - \frac{53}{33} E } .
 \ee

 \noindent ii) for $\frac{15}{106} |\Lambda | < E < \frac{3}{8} |\Lambda |$,
both $\alpha_{+}$ and $\alpha_{-}$ are negative. In this case, the
perturbation is exponentially decreasing.

 \noindent Thus the dilaton can be stabilized as far as the parameter $E$ is in the
range $0 < E < \frac{3}{8}|\Lambda| $.

\section{Dynamical stabilization}

We have seen that both the radion and the dilaton are stabilized
separately assuming one of the two is fixed. In this section, we
consider the dynamical stabilization when both moduli are
perturbed from their critical values. When the sign of the
parameter $E$ is positive, the volume of the three-dimensional
subspace $\zeta$ will increase indefinitely. Due to this inflation
in the three dimensions, terms proportional to $1/\zeta$ can be
ignored soon after the onset of this regime. We also ignore the
damping term because the presence of this term always contributes
to the stabilization positively. We start our perturbative
analysis from setting the asymptotic values of the two moduli
equal to their critical values. We will examine what happens when
the two fields are deviated from their critical values, $\phi
=\phi_0$ and $\xi =\xi_0$.

Substituting
 \be
 \phi = \phi_0 + \delta \phi , ~~~~ \xi = \xi_0 + \delta \xi ,
 \ee
and keeping only linear terms in the perturbation, we have the
following system of equations for the radion and the dilaton
 \be
 {\ddot {\delta \xi} } + \Bigl(\frac{3}{8}|\Lambda| - E
 + \frac{33}{16} Q^2 \frac{e^{-\frac{1}{2} \phi_0} } {\xi_0^2} \Bigr) {\delta \xi}
 + \frac{33}{32} Q^2 \frac{e^{-\frac{1}{2} \phi_0} } {\xi_0} {\delta \phi} = 0,
 \label{dyndelxieq}
 \ee
 \be
 {\ddot {\delta \phi} }
 + \frac{1}{2} Q^2 \frac{e^{-\frac{1}{2} \phi_0} } {\xi_0^3} {\delta \xi}
 + \frac{1}{8} Q^2 \frac{e^{-\frac{1}{2} \phi_0} } {\xi_0^2} {\delta \phi} = 0.
 \label{dyndelphieq}
 \ee
To find the eigenvalues of the coupled equations, we make the
ansatz:
 \be
 \begin{pmatrix}
 \delta \xi \cr
 \delta \phi \cr
 \end{pmatrix}
 ~=~
 \begin{pmatrix}
 \delta \xi_0 \cr
 \delta \phi_0 \cr
 \end{pmatrix} e^{i \omega t}.     \label{eigenvector}
 \ee
Then the system of equations can be written in matrix form using
(\ref{ximin}),
 \be
 \begin{pmatrix}
 -\omega^2 + 2(\frac{3}{8}|\Lambda| - E) &
  \frac{1}{2} (\frac{3}{8}|\Lambda| - E) {\xi_0}
 \cr
 \frac{8}{33} (\frac{3}{8}|\Lambda| - E) \frac{1}{\xi_0} &
 - \omega^2 +  \frac{2}{33} (\frac{3}{8}|\Lambda| - E)
 \cr
 \end{pmatrix}
 \begin{pmatrix}
 \delta \xi_0 \cr
 \delta \phi_0 \cr
 \end{pmatrix} e^{i \omega t} = 0 .  \label{eveqn}
 \ee

The eigenvalues of $\omega$ are given by
 \be
 \omega^2 \Bigl\{ \omega^2
 - \frac{68}{33} \Bigl(\frac{3}{8}|\Lambda| - E \Bigr) \Bigr\}
 = 0.  \label{omegaeqn}
 \ee
The stability condition for the moduli is that $\omega^2$  have
real and positive roots. The existence of any imaginary part in
$\omega$ means that $e^{ i \omega t}$ can grow exponentially,
which means the instability. Obviously equation (\ref{omegaeqn})
satisfies this stability condition and the nontrivial solution of
$\omega$ are
 \be
 \omega = \pm \sqrt{  \frac{68}{33} \Bigl(\frac{3}{8}|\Lambda| - E \Bigr) }.
 \ee
 In conclusion, the radion and the dilaton can be stabilized dynamically
 to their attractor values.

\section{Conclusions}

We have shown that the radion and the dilaton can be stabilized
simultaneously. Our derivation is based on the dependence of
dilaton, radion, and three-dimensional scale factor on their
effective potentials generated by the brane gas and the RR flux.
The existence of a $(D-3)$-brane gas wrapping the extra dimensions
and a four-form RR flux in the unwrapped dimensions can cause
attractor potentials for dilaton and radion. In order to stabilize
the moduli, one requires both the negative and the positive
sources of forces. The brane gas is the origin of a potential term
that can make the radion to contract. The role of the RR flux is
to prevent the radion from collapse. It gives a logarithmic bounce
for small values of the radion. The dilaton field shows a
confining potential consisting of a sum of positive and negative
exponential walls originated from the flux and the brane gas.

In general, the two criteria for the stabilization of a scalar
field are as follows. First, as a function of scalar field,
$V_{\rm eff} (\phi)$ must have a critical point
$\frac{\partial}{\partial \phi} V_{\rm eff} (\phi_0) = 0$. Second,
the second derivative of the effective potential at the critical
point $\frac{\partial^2}{\partial \phi^2} V_{\rm eff} (\phi_0) $
must be positive. One can easily check that, from
(\ref{dilatoneff}) and (\ref{radioneff}), both conditions are
satisfied when one of the moduli is fixed. Since the dilaton field
and the radion field are coupled, we perturbed both fields from
their critical values and found that they are stable. It is
important that the effective potential for the dilaton comes from
their coupling to gauge field and brane gas. For any models where
the dilaton potential is known in the string theory background,
one can check the condition for stability by the above-mentioned
criteria in the Einstein frame.

One important point in moduli stabilization is the frame. To study
the field equations we worked in the ten-dimensional Einstein
frame. One may study the moduli stabilization in four dimensional
Einstein frame after dimensionally reducing the ten-dimensional
theory to four dimensional theory. For fixed dilaton the two
results should be the same. However, when one tries to fix both
the dilaton and the radion dynamically, one should work in the
ten-dimensional Einstein frame. Analysis based on lower
dimensional effective action may be different from the analysis in
the original higher dimensional action.

In our calculation, we introduced the cosmological constant
without further explanation of its origin. Let us briefly sketch
whether it is possible to achieve the stabilization without this
term. Neglecting the damping terms, since the damping terms
contribute stabilization positively, we consider the time
evolution of the differential equation of the form ${\ddot x} = k
x$. The solution of $x(t)$ is exponential for positive $k$ and
oscillating for negative $k$. The initial parameter condition for
the inflation of three-dimensional scale factor and stabilization
of dilaton and radion can be obtained from Eqs (\ref{eoma}),
(\ref{eomdb}), and (\ref{eomdilaton}), for $D=9$,
 \be
  T_{6} \frac{e^{\frac{3}{4} \phi} }{a^3} >
 3 Q^2 \frac{e^{-\frac{1}{2} \phi} } {b^{12}} .
 \ee
Naively it seem that our stabilization argument can be applied to
the cases without cosmological constant term for some initial
values of the moduli. More studies regarding the cosmological
constant are needed.

\begin{acknowledgments}
We would like to thank Robert Brandenberger for suggestions and
comments and the organizers of ``CosPA 2008'' at APCTP, Pohang,
where the idea was initiated. This work was supported by Korea
Research Foundation Grant No. 2009-0070658.
\end{acknowledgments}


\begin{thebibliography}{99}

\bibitem{silverstein}
E. Silverstein, hep-th/0405068; L. McAllister and E. Silverstein,
Gen. Rel. Grav. {\bf 40}, 565 (2008).

\bibitem{bv}
R. Brandenberger and C. Vafa, Nucl. Phys. B {\bf 316}, 391 (1989);
A. A. Tseytlin and C. Vafa, Nucl. Phys. B {\bf 372}, 443 (1992);
A. A. Tseytlin, Class. Quant. Grav. {\bf 9}, 979 (1992).

\bibitem{abe}
S. Alexander, R. Brandenberger, and D. Easson, Phys. Rev. D {\bf
62}, 103509 (2000).

\bibitem{wb0307044}
S. Watson and R. Brandenberger, J. Cocmol. Astropart. Phys.
 {\bf 0311}, 008 (2003).

\bibitem{pb}
 S. P. Patil and R. Brandenberger, Phys. Rev. D {\bf 71}, 103522 (2005);
J. Cocmol. Astropart. Phys. {\bf 0601}, 005 (2006).

\bibitem{bw0403075}
T. Battefeld and S. Watson, J. Cocmol. Astropart. Phys. {\bf
0406}, 001 (2004).

\bibitem{watson}
S. Watson, Phys. Rev. D {\bf 70}, 066005 (2004).

\bibitem{cwb}
Y-K. E. Cheung, S. Watson, and R. Brandenberger, J. High Energy
Phys. {\bf 0605}, 025 (2006).

\bibitem{campos}
A. Campos, Phys. Rev. D {\bf 71}, 083510 (2005).

\bibitem{rador0504047}
T. Rador, Eur. Phys. J. C {\bf 49}, 1083 (2007).

\bibitem{patil}
S. P. Patil, hep-th/0504145.

\bibitem{kaya}
A. Kaya, Phys. Rev. D {\bf 72}, 066006 (2005).

\bibitem{eastro}
D. A. Easson and M. Trodden, Phys. Rev. D {\bf 72}, 026002 (2005).

\bibitem{chatrabhuti}
A. Chatrabhuti, Int. J. Mod. Phys. A {\bf 22}, 165 (2007).

\bibitem{kim0608131}
J. Y. Kim, Phys. Lett. B {\bf 652}, 43 (2007).

\bibitem{akk}
S. Arapoglu, A. Karakci, and A. Kaya, Phys. Lett. B {\bf 645}, 255
(2007).

\bibitem{kim08040073}
J. Y. Kim, Phys. Rev. D {\bf 78}, 066003 (2008).

\bibitem{bc}
A. J. Berndsen and J. M. Cline, Int. J. Mod. Phys. A {\bf 19},
5311 (2004).

\bibitem{bbc}
A. Berndsen, T. Biswas, and J. M. Cline, J. Cocmol. Astropart.
Phys. {\bf 0508}, 012 (2005).

\bibitem{bbem}
T. Biswas, R. Brandenberger, D. A. Easson, and A. Mazumdar, Phys.
Rev. D {\bf 71}, 083514 (2005).

\bibitem{ks}
S. Kanno and J. Soda, Phys. Rev. D {\bf 72}, 104023 (2005).

\bibitem{cw}
S. Cremonini and S. Watson, Phys. Rev. D {\bf 73}, 086007 (2006).

\bibitem{rador0701029}
T. Rador, Eur. Phys. J. C {\bf 52}, 683 (2007).

\bibitem{sano}
M. Sano and H. Suzuki, Phys. Rev. D {\bf 78}, 064045 (2008);
arXiv:0907.2495 [hep-th].

\bibitem{dfb}
R. J. Danos, A. R. Frey, and R. H. Brandenberger, Phys. Rev. D
{\bf 77}, 126009 (2008).


\bibitem{drs}
K. Dasgupta, G. Rejesh, and S. Sethi, J. High Energy Phys. {\bf
9908}, 023 (1999).

\bibitem{gkp}
S. B. Giddings, S. Kachru, and J. Polchinski, Phys. Rev. D {\bf
66},106006 (2002).

\bibitem{gw}
W. D. Goldberger and M. B. Wise, Phys. Rev. Lett. {\bf 83}, 4922
(1999).




\end{thebibliography}
\end{document}